\documentclass[prd,aps,nofootinbib,superscriptaddress,showpacs,floatfix,preprintnumbers,twocolumn]{revtex4}
\usepackage{amssymb,amsmath,graphicx,xcolor,slashed,float,mathrsfs}
\usepackage[colorlinks=true, linkcolor=blue, urlcolor=blue, citecolor=blue]{hyperref}

\begin{document}

\title{Kaon Gluon Parton Distribution and Momentum Fraction from 2+1+1 Lattice-QCD with High Statistics}

\author{Alex NieMiera}
\email{niemiera@msu.edu}
\affiliation{Department of Physics and Astronomy, Michigan State University, East Lansing, Michigan 48824, USA}
\affiliation{Department of Computational Mathematics,
  Science and Engineering, Michigan State University, East Lansing, Michigan 48824, USA}

\author{William Good}
\affiliation{Department of Physics and Astronomy, Michigan State University, East Lansing, Michigan 48824, USA}
\affiliation{Department of Computational Mathematics,
  Science and Engineering, Michigan State University, East Lansing, Michigan 48824, USA}

\author{Huey-Wen Lin}
\affiliation{Department of Physics and Astronomy, Michigan State University, East Lansing, Michigan 48824, USA}

\preprint{MSUHEP-25-025}
\begin{abstract}
We present a high-statistics lattice-QCD determination of the kaon gluon parton distribution function and gluon momentum fraction.
We use clover valence fermion action to take 1,296,640 kaon-correlator measurements on a HISQ ensemble with $a \approx 0.12$~fm and 310-MeV pion mass, generated by the MILC collaboration.
A detailed investigation into the impact of gauge-link smearing on the gluonic matrix elements indicates that five steps of hypercubic smearing offer an effective balance between signal quality and preservation of long-distance physics.
We report a nonperturbatively renormalized kaon gluon momentum fraction of $\langle x \rangle_g^{\overline{\text{MS}}, K} = 0.557(18)_\text{stat}(24)_\text{NPR}(56)_\text{mixing}$ at $\mu = 2$~GeV in in the $\overline{\text{MS}}$ scheme.
Using reduced pseudo-ITD matrix elements and pseudo-PDF matching, we extract the kaon gluon PDF and compare with the prediction from the Dyson-Schwinger equation and with the pion PDF obtained from the same ensemble.
\end{abstract}

\maketitle

\section{Introduction}

Studying kaon structure is crucial for deepening our understanding of the mechanisms behind emergent hadronic mass and the interplay between QCD dynamics and the Higgs mechanism.
Unlike pions, kaons contain a heavier strange quark, making them an ideal system to explore how Higgs-driven mass generation modulates emergent mass mechanisms.
Approximately 20\% of the kaon's mass can be attributed directly to the Higgs mechanism, offering unique insights into the balance of quark and gluon energy contributions to hadron masses.
Furthermore, kaons play a vital role in probing the differences between valence-quark, sea-quark, and gluon distributions compared to pions, shedding light on the universality of QCD dynamics across different mesons.
Understanding kaon structure also has broader implications for mapping the distributions of charge, mass, and spin within hadrons.
The parton distribution functions (PDFs) of the kaon remain less explored compared to those of the pion, with experimental data being particularly sparse.
The gluon PDF in kaons is especially important to study, because gluons play a central role in emergent hadronic mass mechanisms and contribute significantly to the kaon's structure.
Future precision measurements of the kaon gluon PDF at the Electron-Ion Collider (EIC) will provide critical insights into the role of gluons in shaping the properties of mesons~\cite{Achenbach:2023pba,Arrington:2021biu,Aguilar:2019teb}.

Lattice QCD provides a powerful ab-initio framework to calculate kaon PDFs, offering predictions for valence, sea, and gluon distributions.
In the last decade or so, new methods, such as large momentum effective theory (LaMET)~\cite{Ji:2013dva} and pseudo-PDF method~\cite{Balitsky:2019krf}, have been widely used to study the $x$-dependence of kaon distributions.
The first $x$-dependent calculation of kaon structure was of its distribution amplitude (DA)~\cite{Zhang:2017zfe}, an important universal input to hard exclusive processes and form factors at large momentum transfer;
since then, there have been more followup studies~\cite{Zhang:2020gaj,LatticeParton:2022zqc,Cloet:2024vbv} with the continuum-limit taken and with some calculations directly at physical pion mass.
There has been also an exploratory calculation of the kaon $x$-dependent valence-quark PDFs using LaMET~\cite{Lin:2020ssv}.
This study found the ratio of the $u$ quark PDF in the kaon to that in the pion agrees with the CERN NA3 experiment and predictions of the strange-quark distribution of the kaon was made.

In contrast, the gluon structure of the kaon is rarely studied.
Even with the traditional moment operator product expansion (OPE), there has been only one group studying the koan valence-quark lowest 3 moments by ETMC~\cite{Alexandrou:2020gxs,Alexandrou:2021mmi}, and recently gluon momentum fraction of koan~\cite{ExtendedTwistedMass:2024kjf}.
MSULat group presented the first lattice QCD calculation of the kaon's gluon PDF using the pseudo-PDF approach using $2+1+1$-flavor highly improved staggered quark (HISQ) ensembles with pion masses around 310~MeV and lattice spacings of 0.15 and 0.12~fm~\cite{Salas-Chavira:2021wui} with boosted the kaon to momenta around 2~GeV and statistics of 324,000 measurements.
The extracted gluon PDF, determined in the $\overline{\text{MS}}$ scheme at a scale of 2~GeV, showed consistency with phenomenological models, particularly at the 0.12-fm lattice spacing.
Although the mixing between gluon and singlet-quark sectors has been neglected, the effect is expected to be smaller than the gluon signal-to-noise statistical error.

This work extends the previous calculations done by MSULat group, increasing the total number of kaon-correlator measurements to 1,296,640 on the ensemble with smaller lattice spacing.
The rest of the paper is organized as follows.
In Sec.~\ref{sec:LQCD-MEs}, we cover the details of the lattice calculation and the gluon operator used, and discuss the strategy used to extract the bare kaon ground-state matrix elements on the lattice.
In Sec.~\ref{sec:results}, we outline our approach for extracting the kaon gluon momentum fraction and PDF from the lattice data and present our findings.
Finally, we conclude and consider future prospects for improving the existing calculation in Sec.~\ref{sec:conclusion}.

\section{Lattice Matrix Elements and Smearing Study}
\label{sec:LQCD-MEs}

For this work, we use one ensemble of $N_f = 2 + 1 + 1$ HISQ~\cite{Follana:2006rc}, generated by the MILC Collaboration~\cite{MILC:2012znn}, with a lattice spacing of $a = 0.1207(11)$~fm and a valence pion mass of $M_\pi^\text{val} = 0.309(1)$~GeV.
We use Wilson-clover fermions in the valence sector and tune the valence quark masses to reproduce the mass of the lightest meson containing light and strange quarks.
The calculation is carried out across 1013 gauge-field configurations, with a total of 1,296,640 two-point (2pt) correlation function measurements.
The two-point correlation functions are defined by
\begin{equation} \label{eq:2pt-correlator}
    C^{\text{2pt}} (P_z; t) = \int d^3 y \; e^{-i y P_z} \langle 0 | \chi (\vec{y}, t) | \chi (\vec{0},0) | 0 \rangle,
\end{equation}
where $t$ is Euclidean time, $P_z$ is the boost momentum in the z-direction, and $\chi = \bar{q}_1 \gamma_5\gamma_t q_2$ is an interpolating operator, which we find to have good overlap with the ground state at both the zero and nonzero momenta with $q_i$ to be strange and up/down quarks.

The three-point (3pt) correlators, which we construct by contracting our two-point correlators with gluon loops, are defined by
\begin{multline} \label{eq:3pt-correlator}
    C^{\text{3pt}}(P_z,  z; t_\text{sep}, t) = \\
    \int d^3 \; y e^{-iyP_z} \langle 0 | \chi (\vec{y}, t_\text{sep}) O_g(z,t) \chi(\vec{0}, 0) | 0 \rangle,
\end{multline}
where $t_\text{sep}$ is the separation time between the source and sink, and $t$ is the insertion time of a gluon operator, $O_g (z,t)$.
To obtain the gluon PDF, we use the the unpolarized gluon PDF operator first discussed in Ref.~\cite{Balitsky:2019krf}
\begin{multline} \label{eq:Balitsky-gluon-operator}
    O^\text{RpITD}_g(z) = \\
    F^{ti}(z)W(z,0)F_i^t(0) - F^{ij}(z) W(z,0) F_{ij}(0),
\end{multline}
where $i,j$ denote the summation over transverse indices $\{x,y\}$, $F_a^{\mu \alpha} = \partial^\mu A_a^\alpha - \partial^\alpha A_a^\mu - g f_{abc} A_b^\mu A_c^\alpha$ is the gluon field strength tensor, and the gauge-invariant Wilson line is
\begin{equation} \label{eq:wilson-line}
    W(z,0) = \mathcal{P} \exp \left[ -i g \int_0^z dz' \; A^z(z')\right],
\end{equation}
with $A^z = A_a^z t_a$.
In this work, we also report the kaon gluon momentum fraction with the gluon moment operator
\begin{equation} \label{eq:moment-gluon-operator}
    O_g^\text{OPE} = F^{t \mu} F^{\mu}_t - \frac{1}{4} F^{\mu \nu} F_{\mu \nu},
\end{equation}
following the recent lattice studies~\cite{Shanahan:2018pib,ExtendedTwistedMass:2021rdx,Fan:2022qve,Hackett:2023nkr,Good:2023ecp,ExtendedTwistedMass:2024kjf}.

To improve the signal of the noisy gluon matrix elements, gauge-link smearing is commonly employed to reduce short-distance fluctuations while preserving long-range physics.
The smearing techniques used in, for example, recent nucleon gluon PDF studies from different groups~\cite{HadStruc:2021wmh,Fan:2022kcb,Delmar:2023agv} include Wilson flow~\cite{Luscher:2010iy}, hypercubic (HYP) smearing~\cite{Hasenfratz:2001hp}, and Stout smearing~\cite{Morningstar:2003gk}.
In a previous unpublished study~\cite{Good:2023rbp}, we analyzed the effects of gauge-link smearing across these techniques and numbers of smearing steps for pseudo-PDF matrix elements of heavy nucleons and pions.

In this work, we also perform a gauge-smearing study for the kaon pseudo-PDF matrix elements using the previously mentioned smearing techniques.
We apply the Wilson-flow smearing~\cite{Luscher:2010iy}, whose gauge link varies with ``flow time'' $t$, starting from $V_{i,\mu;t=0} = U_{i,\mu}$ and updated by the differential equation
\begin{equation}
    \dot{V}_{i,\mu;t} = -g_0^2 (\left\{\partial_{x,\mu}S_\text{w}(V_{i,\mu;t})\right\}V_{i,\mu;t}),
\end{equation}
where $S_\text{w}(U)$ is the Wilson action and $g_0$ is the bare coupling.
We use $N_\text{steps} = 100$ steps for the Wilson flow times $t = a^2$ and $3a^2$, which we label ``WILSON1'' and ``WILSON3'', both within the ranges explored and used to extrapolate back to zero flow time in Ref.~\cite{HadStruc:2021wmh} for the nucleon gluon matrix elements.
We also adopt HYP smearing~\cite{Hasenfratz:2001hp} which consists of 3 stages, where each stage creates a composite object of gauge links with one less Lorentz index:
\begin{subequations}
\label{eq:HYP-Smearing}
\begin{multline}
    \bar{V}_{i,\mu; \nu \, \rho} = P_\text{SU(3)} \Bigg[(1-\alpha_{3})U_{i,\mu} + \\
    \frac{\alpha_{3}}{2} \sum_{\pm \eta \neq \rho, \nu, \mu} U_{i,\eta} U_{i+\hat{\eta},\mu} U_{i+\hat{\mu},\eta}^{\dagger}\Bigg],
\end{multline}
\begin{multline}
\tilde{V}_{i,\mu; \nu}= P_\text{SU(3)}\Bigg[(1-\alpha_{2})U_{i,\mu}+ \\
\frac{\alpha_{2}}{4}\sum_{\pm \rho \neq \nu, \mu}\bar{V}_{i,\rho; \nu \, \mu}\bar{V}_{i+\hat{\rho},\mu; \rho \, \nu}\bar{V}_{i+\hat{\mu},\rho; \nu \, \mu}^{\dagger}\Bigg],
\end{multline}
\begin{multline}
V_{i,\mu}= P_\text{SU(3)} \Bigg[(1-\alpha_{1})U_{i,\mu} + \\
\frac{\alpha_{1}}{6} \sum_{\pm \nu \neq \mu} \tilde{V}_{i,\nu; \mu} \tilde{V}_{i+\hat{\nu},\mu; \nu} \tilde{V}_{i+\hat{\mu},\nu; \mu}^{\dagger}\Bigg],
\end{multline}
\end{subequations}
where $P_\text{SU(3)}$ is the projector back onto SU(3) and $V_{i,\mu}$ are the final smeared links.
This creates smeared links, which only have contributions from the hypercubes containing the original thin link.
We use the common choice of HYP-smearing parameters: $\alpha_1 = 0.75$, $\alpha_2 = 0.6$, $\alpha_3 = 0.3$.
The process described in Eq.~\ref{eq:HYP-Smearing} represents one step of HYP smearing, and in this work, we explore applying $X \in \{3,5,7,10\}$ steps of HYP smearing, which we label ``HYP$X$'', where HYP5 has been used in MSULat's previous gluon PDF studies~\cite{Fan:2021bcr,Salas-Chavira:2021wui,Fan:2022kcb,Fan:2022qve,Good:2023ecp,Good:2024iur}.
Finally, we also use Stout smearing~\cite{Morningstar:2003gk} to improve our signal, which is defined by adding a weighted sum of perpendicular staples to the original gauge field:
\begin{multline}
V_{i,\mu} = \\
P_\text{SU(3)} \biggl[U_{i,\mu} + \sum_{\nu\neq \mu} \rho_{\mu\nu} \biggl(U_\nu(x) U_\mu(x+\hat{\nu}) U_\nu^\dagger(x+\hat{\mu}) \\
 + U^\dagger_\nu(x-\hat{\nu}) U_\mu(x-\hat{\nu}) U_\nu(x-\hat{\nu}+\hat{\mu})
\biggr)\biggr] .
\end{multline}
We use $\rho_{jk} = \rho = 0.125$ and $\rho_{4\mu} = \rho_{\mu 4} = 0$.
Similar to HYP smearing, we compare the use of multiple steps, $Y \in \{10,20\}$ of Stout smearing, which denote as ''STOUT$Y$''.
These two values are explored in Ref.~\cite{Delmar:2023agv} for a study of the nucleon gluon PDF.

From the two-point correlators and three-point correlators with various smearing methods and amounts, we may extract bare lattice matrix elements for our operators.
We first fit the two-point correlators using the two-state fit form
\begin{equation} \label{eq:2pt-correlator-two-state}
    C^{\text{2pt}} (P_z; t) = |A_0|^2 e^{-E_0 t} + |A_1|^2 e^{-E_1 t},
\end{equation}
with the ground (excited) state amplitudes and energies respectively given by $A_0$ ($A_1$) and $E_0$ ($E_1$).
We then use these fitted energies and amplitudes to perform a simultaneous fit across multiple separation times to the two-state fit form of the three-point correlators
\begin{align} \label{eq:3pt-correlator-two-state}
    C^{\text{3pt}} (P_z, z; t_\text{sep}, t) &= |A_0|^2 \langle 0 | O_g | 0 \rangle e^{-E_0 t_\text{sep}} \notag \\
    & + |A_0| |A_1| \langle 0 | O_g | 1 \rangle e^{-E_0 (t_\text{sep} - t)} e^{-E_0 t} \notag \\
    & + |A_0| |A_1| \langle 1 | O_g | 0 \rangle e^{-E_1 (t_\text{sep} - t)} e^{-E_0 t} \notag \\
    & + |A_1|^2 \langle 1 | O_g | 1 \rangle e^{-E_1 t_\text{sep}},
\end{align}
with the ground-state, ground–excited--state, and excited-state matrix elements respectively given by $\langle 0 | O_g| 0 \rangle$, $\langle 0 | O_g | 1 \rangle = \langle 1 | O_g | 0 \rangle$, and $\langle 1 | O_g | 1 \rangle$.
We then take the ratio of the fitted two- and three-point correlators to verify the reliability of our extracted ground state matrix elements
\begin{equation} \label{eq:2pt-3pt-correlator-ratio}
    R(t_\text{sep}, t) = \frac{C^{\text{3pt}} (P_z, z; t_\text{sep}, t)}{C^{\text{2pt}} (P_z; t_\text{sep})}.
\end{equation}
Ideally, this ratio asymptotically approaches the ground-state matrix element as $t_\text{sep} \rightarrow \infty$.

\begin{figure*}[p]
    \centering
    \includegraphics[width=0.9\textwidth]{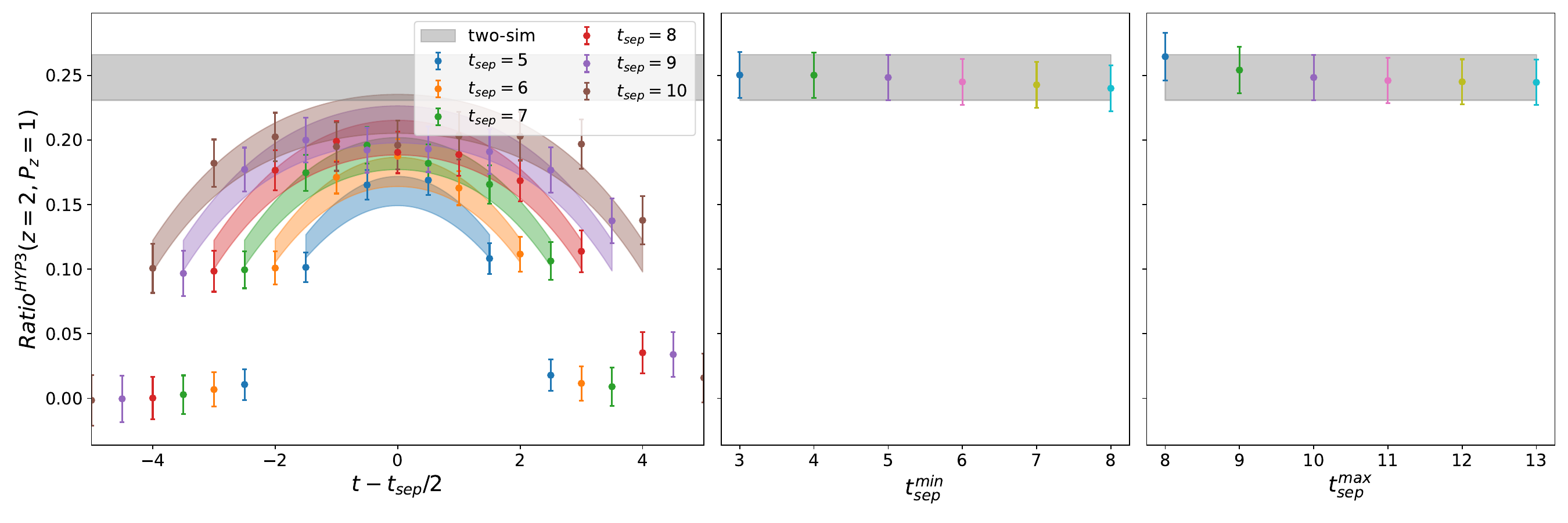}\\
    \includegraphics[width=0.9\textwidth]{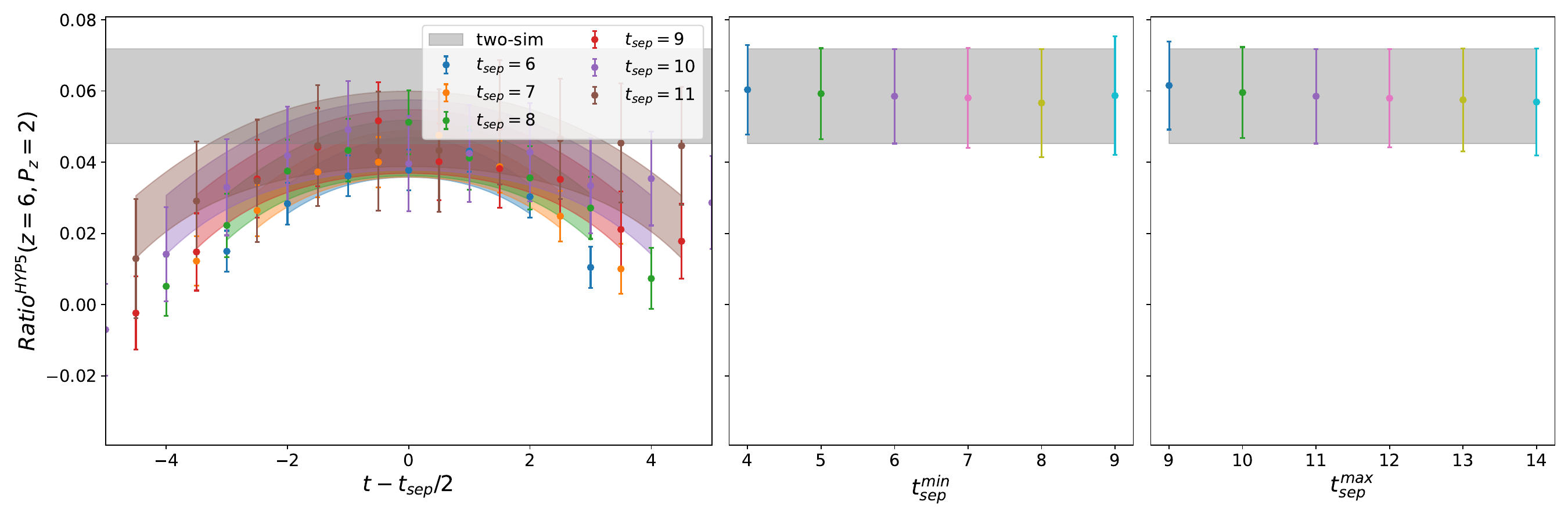}\\
    \includegraphics[width=0.9\textwidth]{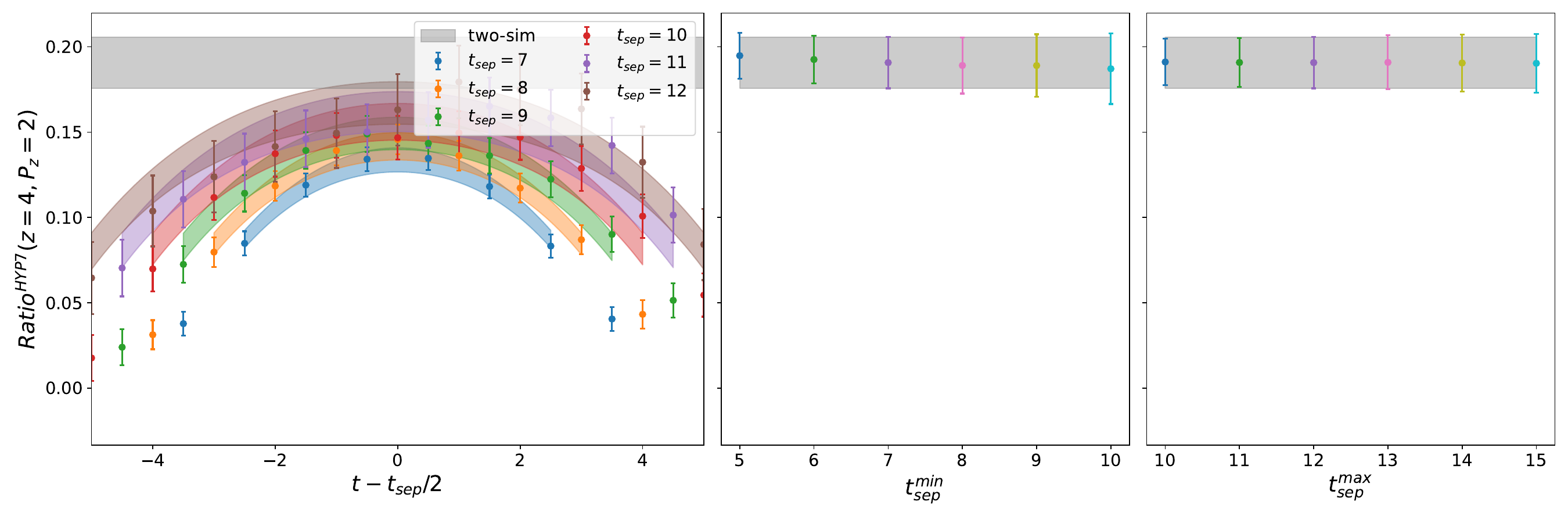}\\
    \includegraphics[width=0.9\textwidth]{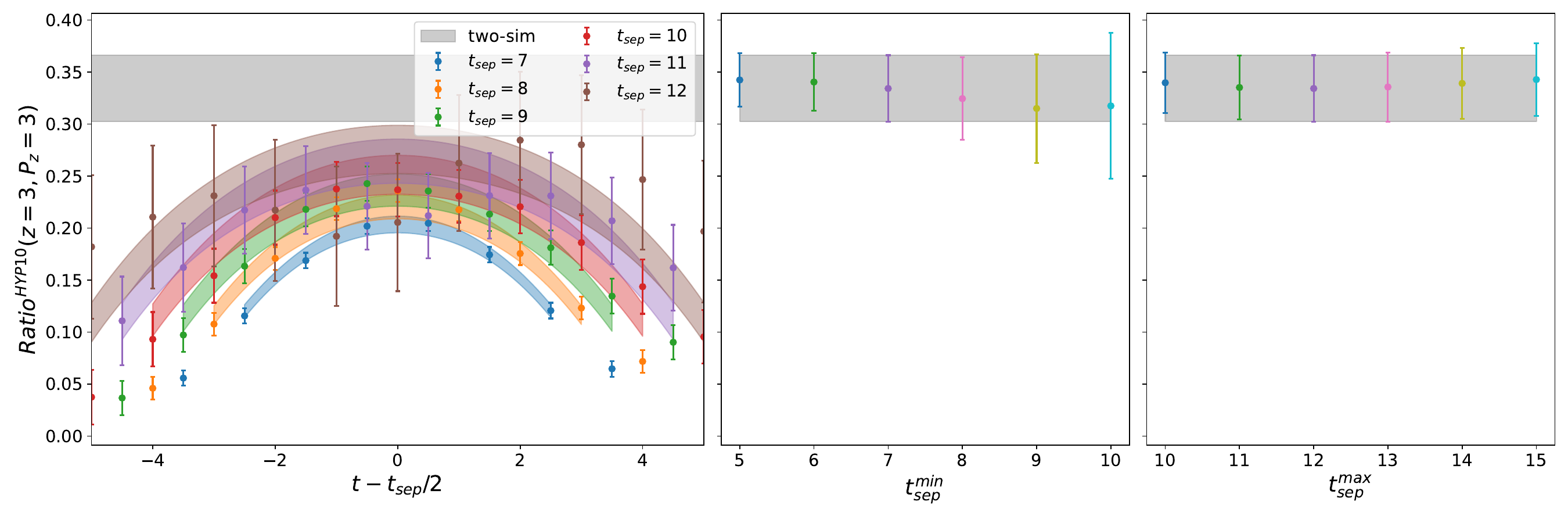}\\
    \caption{
    Example ratio plots used to extract the ground-state matrix element with the gluon operator $O_g^\text{RpITD}$.
    Each row corresponds to a different level of HYP smearing: HYP3 (top), HYP5, HYP7, and HYP10 (bottom), with arbitrary choices of momentum $P_z$ and spatial separation $z$.
    The left column shows the ratio of three-point to two-point correlators as a function of source-sink separation time $t_\text{sep}$, which approaches a plateau corresponding to the ground-state contribution.
    The middle and right columns show how the extracted matrix element depends on the choice of minimum ($t_\text{sep}^\text{min}$) and maximum ($t_\text{sep}^\text{max}$) separation times used in the fit, respectively.
    In all panels, the gray band denotes the ground-state matrix element obtained from a simultaneous two-state fit, and the purple points in the middle and right plots indicate the fit range selected for the final analysis.
}
    \label{fig:hyp-smearing-ratio-plots}
\end{figure*}

Figure~\ref{fig:hyp-smearing-ratio-plots} illustrates the extraction of the ground-state matrix element using the gluon operator $O_g^\text{RpITD}$ defined in Eq.~\ref{eq:Balitsky-gluon-operator}, for varying HYP smearing levels, with each row corresponding to HYP3, HYP5, HYP7, and HYP10 from top to bottom at varying values of $z$ in units of the lattice spacing and $P_z$ in units of $2\pi/L$.
Focusing first on the leftmost panel of each row, we plot the ratio of three-point to two-point correlators as a function of the source-sink separation time $t_\text{sep}$.
The data points are the numerical results from our lattice simulations at discrete values of $t_\text{sep}$, and the corresponding colored bands are the ratio reproduced from the simultaneous two-state fit across multiple $t_\text{sep}$.
The gray horizontal band indicates the ground-state matrix element extracted from the fit.
We see that the fits reproduce the ratios from the data within one standard deviation in most cases, and the ratio from the data and fits approach the ground state matrix element as $t_\text{sep}$ increases, where contributions from excited states are exponentially suppressed.

\begin{figure*}
\centering
\includegraphics[width=0.9\textwidth]{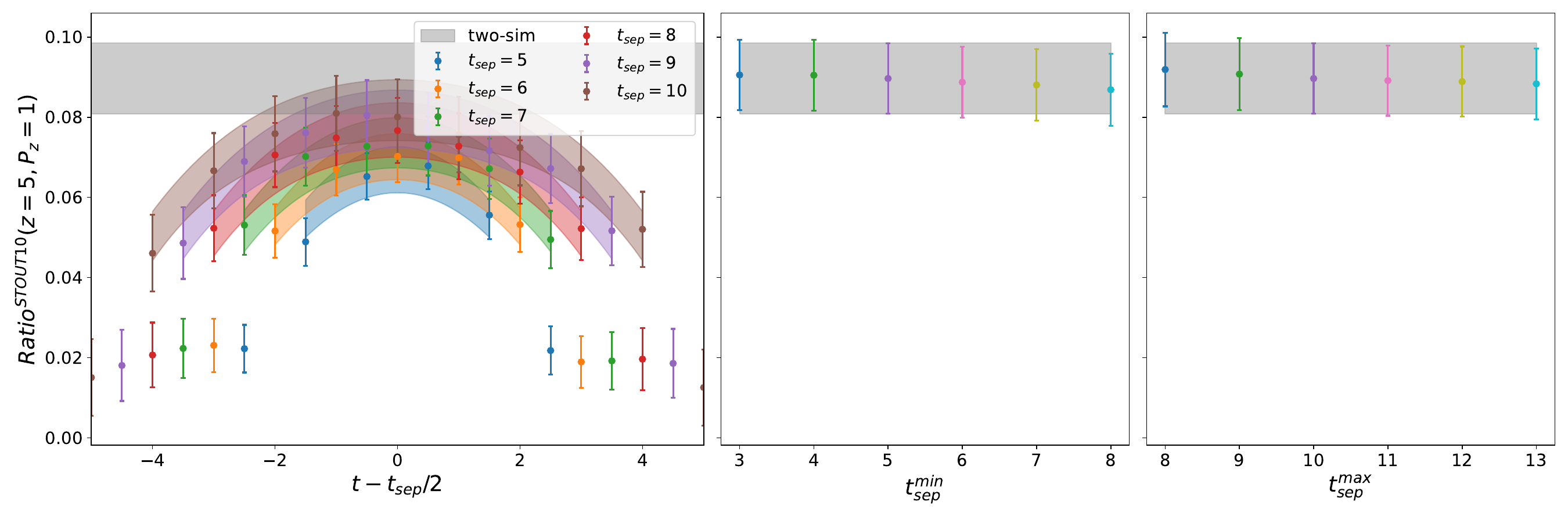}
\includegraphics[width=0.9\textwidth]{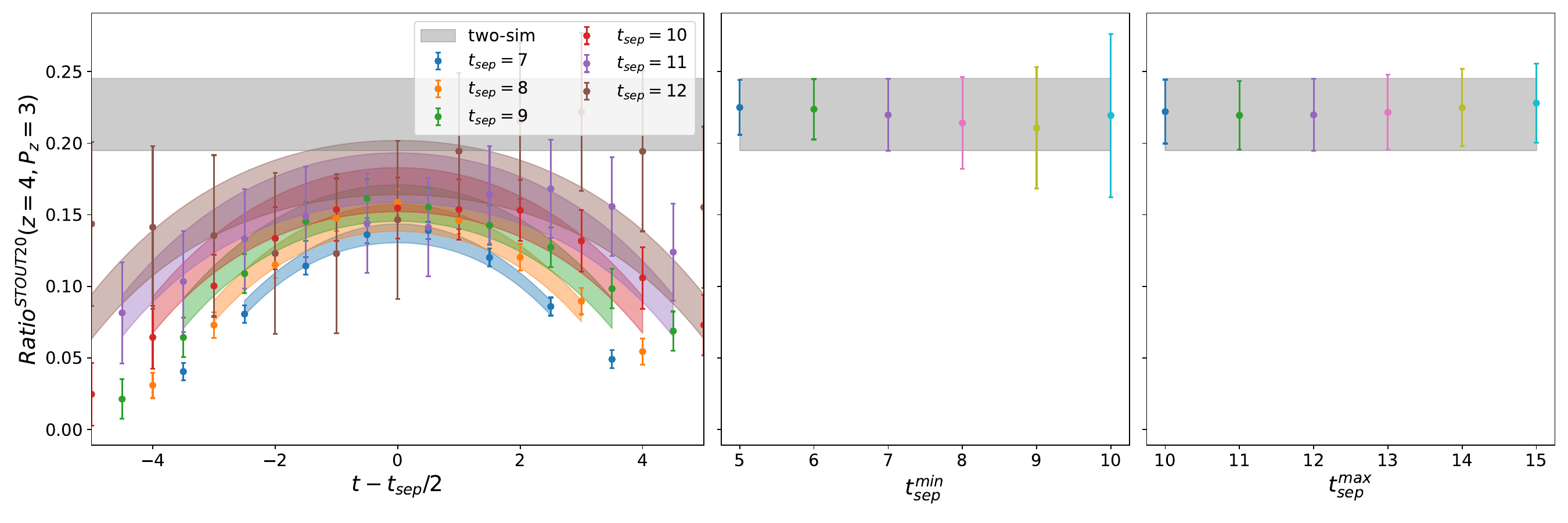}
    \caption{Ratio plots and source-sink separation dependencies for stout smearing with $10$ steps (top row) and $20$ steps (bottom row), using the gluon operator $O_g^\text{RpITD}$.
    Each row follows the same layout as in Fig.~\ref{fig:hyp-smearing-ratio-plots}: the left panel shows the ratio of three-point to two-point correlators as a function of source-sink separation time $t_\text{sep}$, while the middle and right panels display the stability of the extracted matrix element under variations in the minimum and maximum fit ranges, respectively.
    The gray band denotes the final matrix element from the selected fit, and the purple points in the middle and right plots indicate the chosen fit window.}
    \label{fig:stout-smearing-ratio-plots}
\end{figure*}

\begin{figure*}
    \centering
    \includegraphics[width=0.9\textwidth]{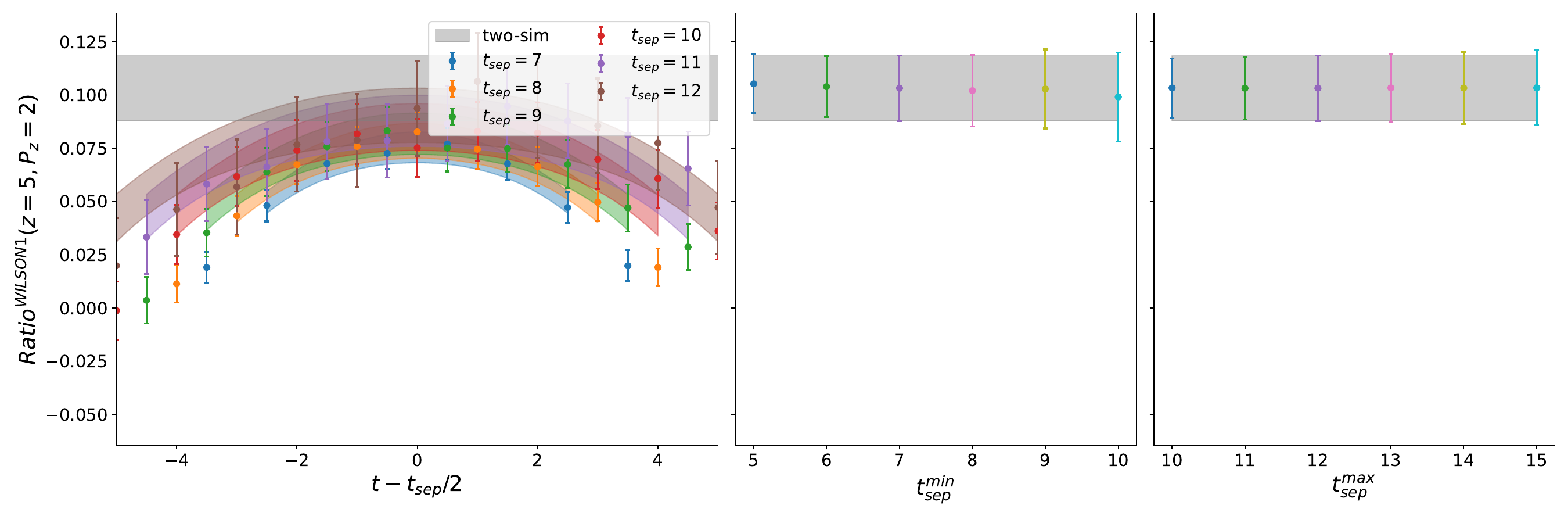}
    \includegraphics[width=0.9\textwidth]{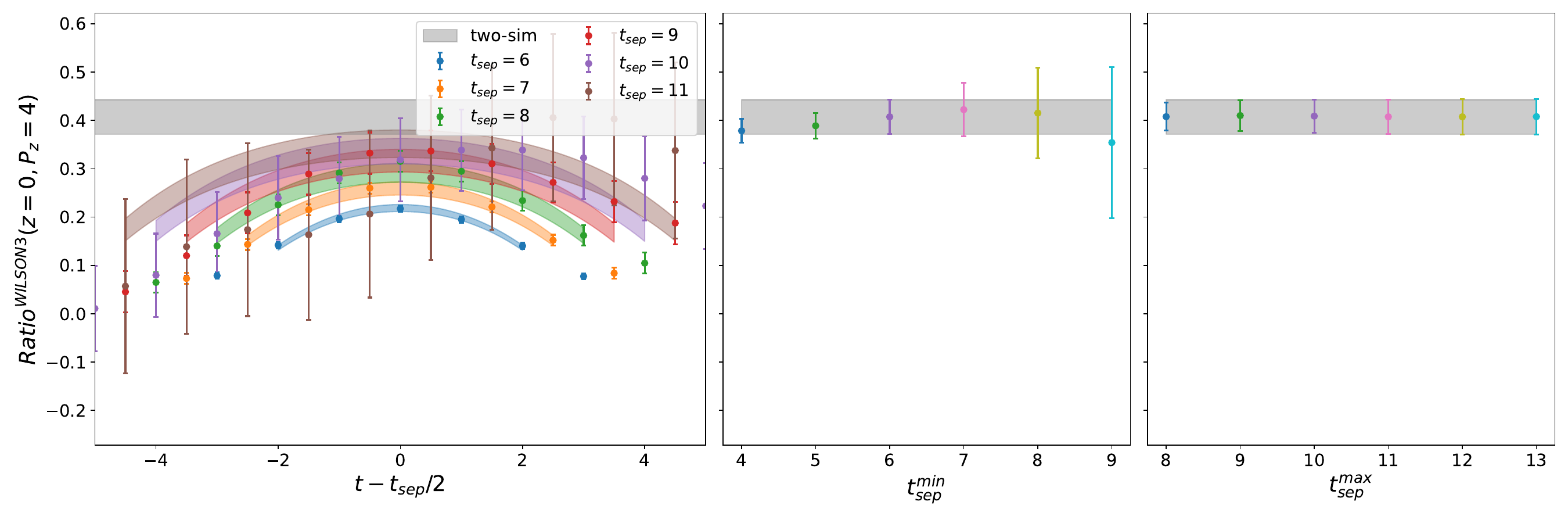}
    \caption{Ratio plots and source-sink separation dependencies for Wilson smearing with $1$ steps (top row) and $3$ steps (bottom row), using the gluon operator $O_g^\text{RpITD}$.
    Each row follows the same layout as in Fig.~\ref{fig:hyp-smearing-ratio-plots}: the left panel shows the ratio of three-point to two-point correlators as a function of source-sink separation time $t_\text{sep}$, while the middle and right panels display the stability of the extracted matrix element under variations in the minimum and maximum fit ranges, respectively.
    The gray band denotes the final matrix element from the selected fit, and the purple points in the middle and right plots indicate the chosen fit window.}
    \label{fig:wilson-smearing-ratio-plots}
\end{figure*}

The middle and right columns assess the sensitivity of the extracted matrix element to variations in the fitting window.
The middle panels fix the maximum separation time and vary the minimum included value $t_\text{sep}^\text{min}$, while the rightmost panels do the opposite, fixing $t_\text{sep}^\text{min}$ and varying $t_\text{sep}^\text{max}$.
In both cases, the extracted matrix elements are plotted as a function of the chosen cut, with the purple point marking the fit selected for our final analysis.
The fits with different $t_\text{sep}^\text{min}$ and $t_\text{sep}^\text{max}$ are all well within one standard deviation of each other, demonstrating robustness of our fits against both short-distance excited-state contamination and large-separation statistical noise.
Figures~\ref{fig:stout-smearing-ratio-plots} and~\ref{fig:wilson-smearing-ratio-plots} follow the same format, where the Stout smearing plots correspond to STOUT$10$ and $20$ steps of smearing from top to bottom, while the Wilson smearing plots reflect WILSON$1$ and $3$ steps of smearing in the same vertical order.
In all cases, the ratio plots exhibit $1\sigma$ agreement between the fits and the data, with clear convergence toward a stable ground-state signal, and minimal sensitivity to variations in the fit window.
The consistency observed across these different smearing schemes supports the robustness and reliability of our procedure for extracting the ground-state matrix element, in general.

\begin{figure}
    \centering
    \includegraphics[width=0.5\textwidth]{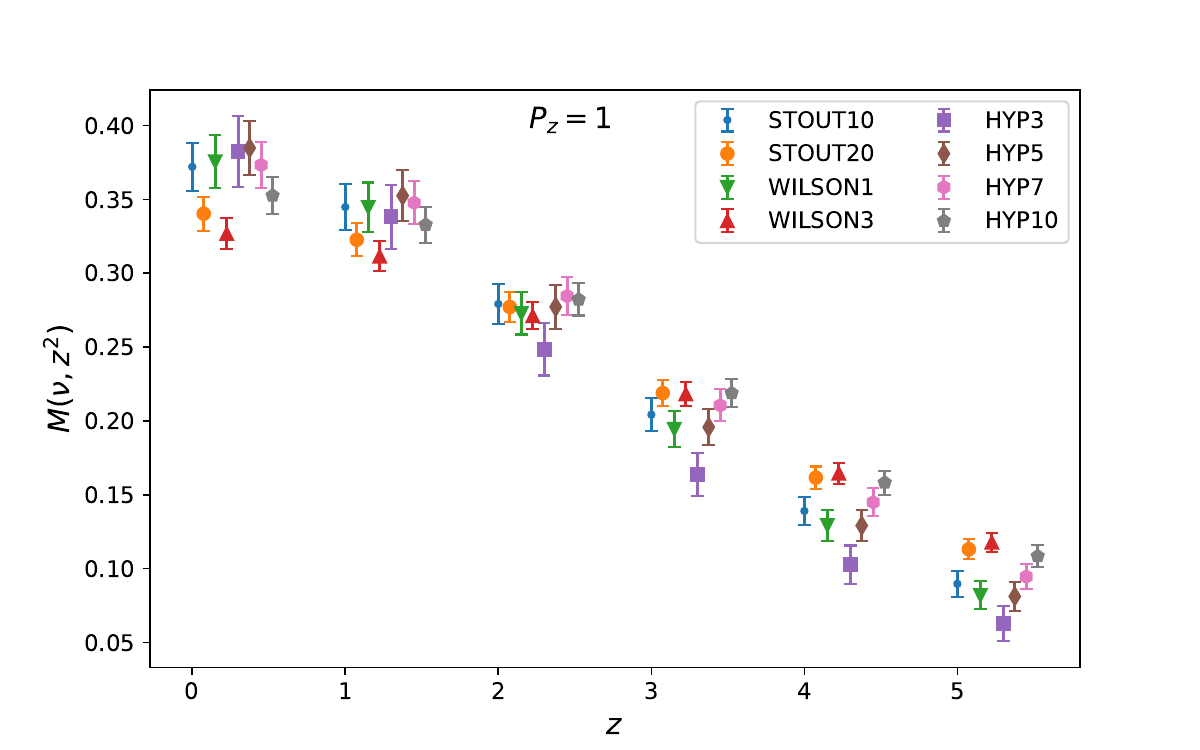}
    \includegraphics[width=0.5\textwidth]{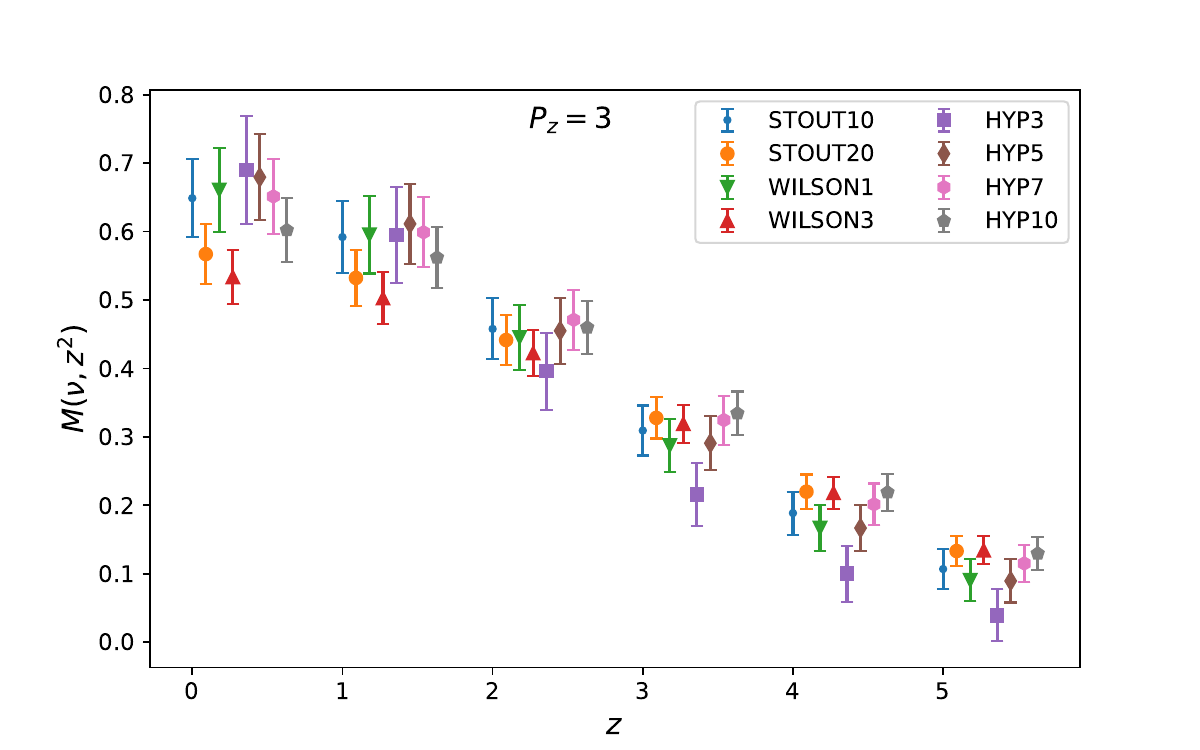}
    \caption{Comparison of the fitted ground state matrix elements with the gluon operator $O_g^\text{RpITD}$ across the considered smearing techniques, plotted against (integer) $z$ for $P_z = 1$ (top) and $P_z = 3$ (bottom).
    The overlapping data points have been horizontally offset for improved visibility, with the blue dots representing the unshifted $z$-value for each cluster of smeared data.}
    \label{fig:GMSE-smearing-study}
\end{figure}

Figure~\ref{fig:GMSE-smearing-study} shows the bare fitted ground-state matrix elements for the gluon operator $O_g^\text{RpITD}$, plotted as a function of Wilson line length $z$, for two different momenta $P_z$.
Each set of points corresponds to a different smearing scheme (HYP, Stout, or Wilson), with multiple smearing levels shown within each group.
For visual clarity, horizontal offsets have been applied to shift the overlapping data points, with the blue points denoting the true, unshifted $z$-position.
While the absolute magnitudes of the matrix elements cannot be directly compared due to differing renormalization factors across smearing types, consistent patterns in their relative sizes still offer some insight into the effect of the different smearing schemes.
For instance, STOUT10 and WILSON1 consistently fall between HYP5 and HYP7 across all values of $z$, suggesting that their effective smearing strengths are comparable.
Additionally, at large $z$, we can see that the signal-to-noise ratio increases with increasing steps of smearing, which is particularly visible in the HYP smeared matrix elements.
These trends are consistent across both momenta and provide some insight into the impact of different smearing techniques on matrix element extraction prior to renormalization.

We compute the reduced pseudo--Ioffe time distribution (RpITD), which cancels renormalization and kinematic factors, reduces lattice systematics, and removes ultraviolet divergences.
We define the RpITD as the double ratio of our fitted matrix elements
\begin{equation} \label{eq:RpITD-double-ratio}
    \mathcal{M}(\nu, z^2) = \frac{M (z \cdot P_z,z^2)/ M(0 \cdot P_z, 0)}{ M(z \cdot 0, z^2) / M(0 \cdot 0, 0)},
\end{equation}
where $M (z \cdot P_z, z^2)$ represents the fitted ground state matrix elements at Wilson line length $z$ and momentum $P_z$ and $\nu = z \cdot P_z$ denotes the Ioffe time.
The cancellation of the renormalization factors and the fact that $\mathcal{M}(\nu = 0, z^2) = 1$ allow for a more meaningful comparison of shapes across different smearing techniques.

\begin{figure}
    \centering
    \includegraphics[width=0.5\textwidth]{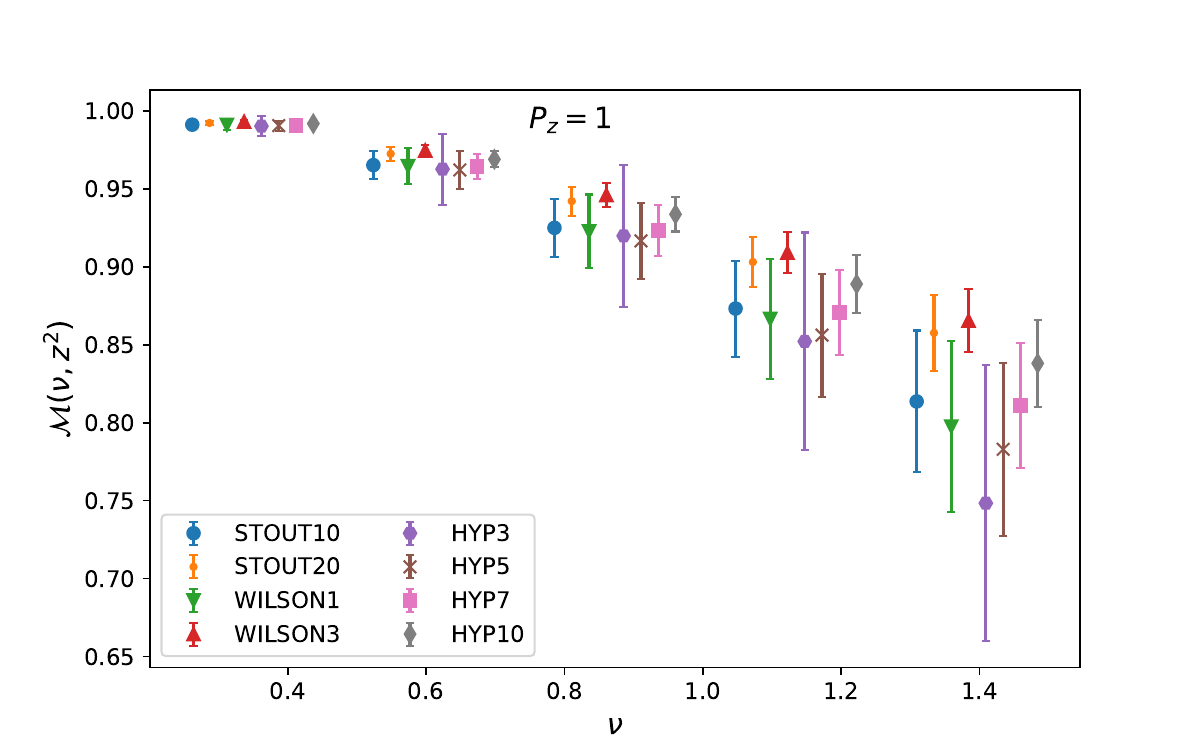}
    \includegraphics[width=0.5\textwidth]{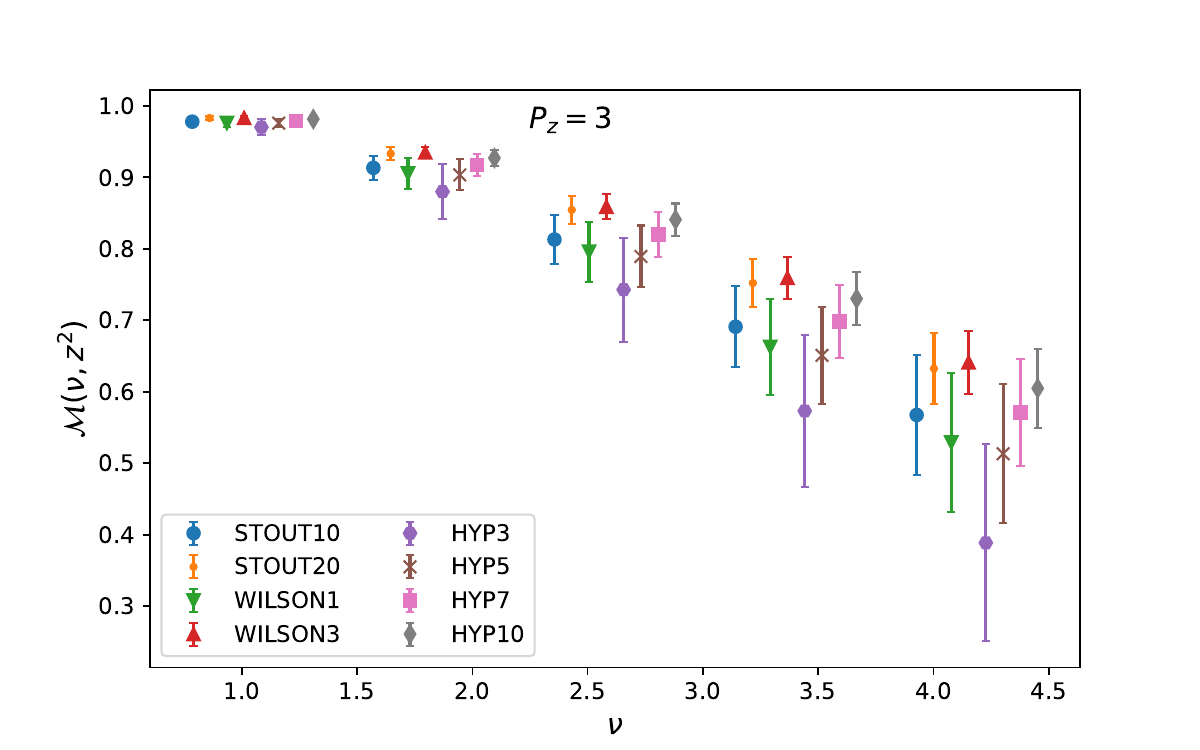}
    \caption{Comparison of the RpITDs computed with the gluon operator $O_g^\text{RpITD}$ for $P_z = 1$ (top) and $P_z = 3$ (bottom), across all considered smearing techniques.
    The overlapping data points have been horizontally offset for improved visibility, with the blue dots representing the unshifted $\nu$-value for each cluster of smeared data.}
    \label{fig:RpITDs-smearing-study}
\end{figure}

In Fig.~\ref{fig:RpITDs-smearing-study}, we present the RpITDs as functions of Ioffe time for two different (integer) momenta $P_z = 1$ and $3$, grouped by smearing technique and step size.
As in the previous plots, the blue points represent the true, unshifted $\nu$-values, serving as visual references for each cluster of data.
While increasing the number of smearing steps in HYP, Stout, and Wilson techniques improves the signal to noise, it also causes the RpITDs to flatten out across all $\nu$, which is expected, as the gluon fields will approach unity, in the limit of large smearing.
This indicates underestimation of the errors and possible loss of physics at the largest smearing.
At this level of statistics, we can see that the largest smearing values, WILSON2 and STOUT20, sometimes fall outside of $1\sigma$ agreement with the lowest HYP smearing values at the largest $\nu$, but even at the most extreme, the agreement is well within $2\sigma$.
Overall, comparing $P_z = 1$ and $3$, it appears that the effect of smearing increases with the invariant Ioffe time.
The RpITDs for HYP5, STOUT10, WILSON1, all have mean values and errors which are quite similar, when compared with the larger smearing data, suggesting that these smearing techniques yield comparable results.
Similarly, HYP10, STOUT20, and WILSON3 also show behavior indicating comparable results.
In the intermediate range, the smearing is strong enough to reduce noise without excessively suppressing the physics;
however, in the more aggressive range, we trade reduced error for a loss of physics.
Using this study of different smearing choices, we find HYP5 to be a conservative choice, which gives a fair estimation of the statistical errors without overly affecting the physics.

\begin{figure*}[h]
    \centering
    \includegraphics[width=0.9\textwidth]{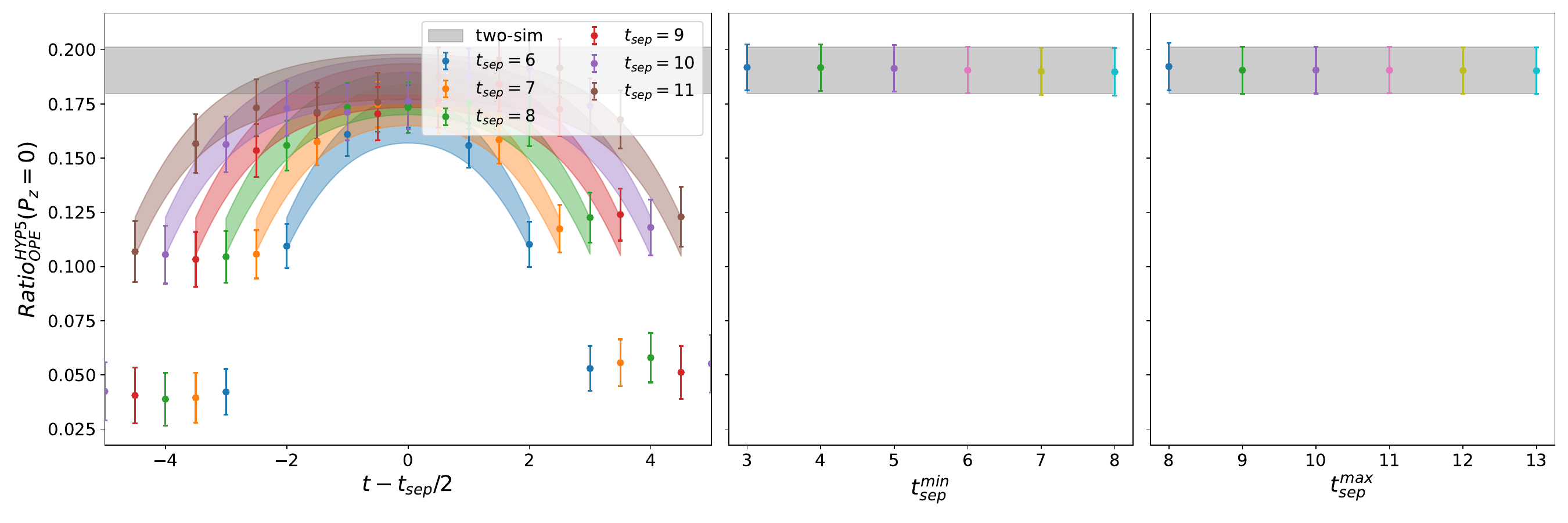}
    \includegraphics[width=0.9\textwidth]{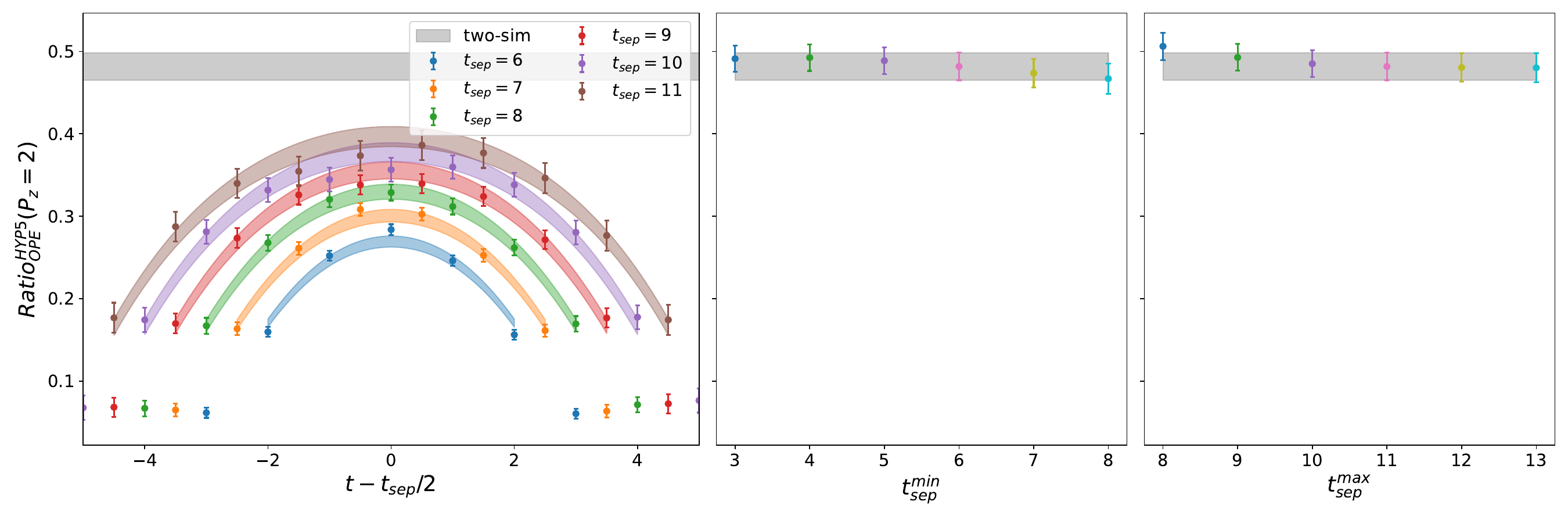}
    \caption{
    Ratio plots and source-sink separation dependencies for HYP$5$ smearing using the gluon operator $O_g^\text{OPE}$, shown for $P_z = 0$ (top row) and $P_z = 2$ (bottom row).
    Each row follows the same layout as in Fig.~\ref{fig:hyp-smearing-ratio-plots}: the left panel displays the ratio of three-point to two-point correlators as a function of the operator insertion time;
    the middle and right panels show the stability of the extracted matrix element with respect to variations in the minimum and maximum fit bounds, respectively.
    The gray band represents the final matrix element from the selected fit, with the pink point marking the preferred fit window.
    }
    \label{fig:HYP5-O0ZF-smearing-ratio-plots}
\end{figure*}

For the gluon moment operator $O_g^\text{OPE}$, the renormalization is more complicated, so we only consider the HYP5 results for this operator.
We follow the same analysis process when extracting the ground-state matrix elements for the gluon moment operator $O_g^\text{OPE}$ and show representative ratio plots in Fig.~\ref{fig:HYP5-O0ZF-smearing-ratio-plots}, evaluated using our final choice of HYP5 smearing across several source-sink separations, which are similar to Figs.~\ref{fig:hyp-smearing-ratio-plots}-\ref{fig:wilson-smearing-ratio-plots}.
The top and bottom panels display results for selected momenta $P_z$, illustrating good agreement between the fits and the data and convergence as the separation time increases.
As in previous ratio plots, the middle and rightmost columns highlight the source-sink dependence, providing a good check on the reliability of the extracted ground-state matrix element.
In these plots, the pink point gives the final fit range $t_\text{sep} \in [6,11]$, which is again, comfortably within $1\sigma$ of all other fit range choices.

\section{Kaon Gluon Moment and PDF Results}
\label{sec:results}
\subsection{Moment Results}
\begin{figure}[h]
\centering
\includegraphics[width=0.45\textwidth]{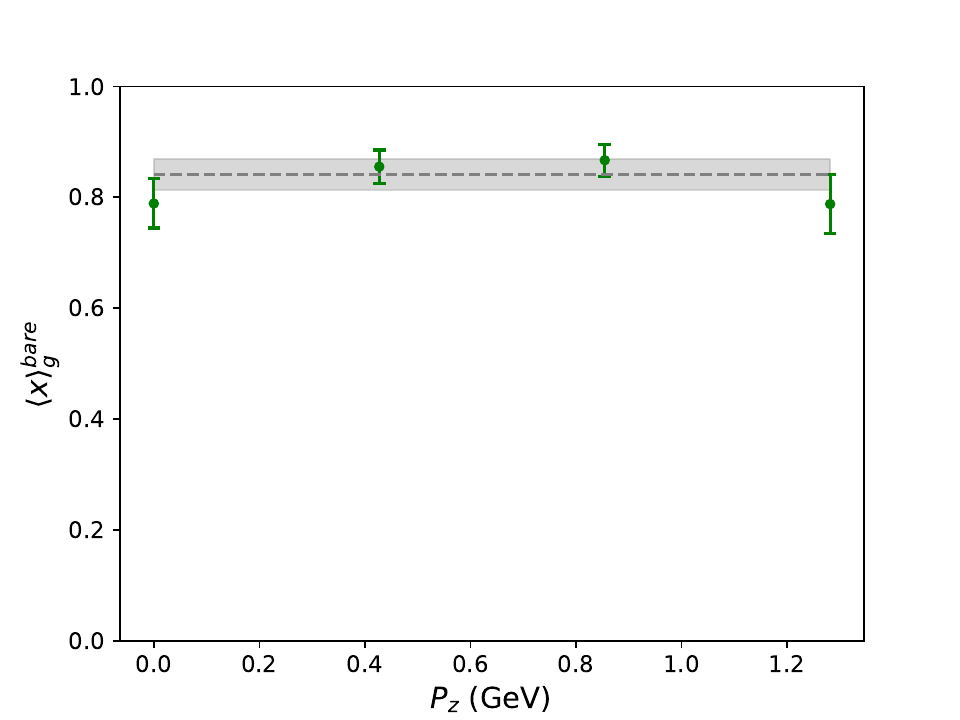}
\caption{ \label{fig:bare-gluon-moment}
    The bare gluon momentum fraction $\langle x \rangle_g^\text{bare}$ (dark green points) as a function $P_z$ in GeV, computed using the gluon operator $O_g^\text{OPE}$ with HYP$5$ smearing.
    The gray band is our final estimate of $\langle x \rangle_g^\text{bare}$ after fitting over all equivalent gluon first moments determined from different boost kaon momenta.
    }
\end{figure}

The bare gluon momentum fraction is related to the extracted ground-state matrix elements of the OPE operator in above section through a kinematic factor
\begin{equation}
   \langle 0 | O_g^\text{OPE} | 0 \rangle = \frac{3E_0^2 + P_z^2}{4 E_0}\langle x \rangle_g^\text{bare},
\end{equation}
where $E_0$ is the ground-state energy extracted from the two-state fit to the two-point kaon correlators using the form defined in Eq.~\ref{eq:2pt-correlator-two-state}.
With this relation, we can compute the bare gluon momentum fraction of kaon at several momenta.
Figure~\ref{fig:bare-gluon-moment} displays the resulting values of $\langle x \rangle_g^\text{bare}$ (dark green points) as a function of $P_z$ in GeV.
We see that the mid-momentum data, $P_z \approx 0.4$ and $0.85$~GeV have the best signal-to-noise ratios due to the momentum smearing used in the calculations.
All of bare momentum fractions calculated at different momenta are consistent with each other within $1\sigma$ of statistical error.
We then take a constant fit over all momenta and come to a final estimate of the bare gluon momentum fraction, also indicated as the gray band in Fig.~\ref{fig:bare-gluon-moment}, $\langle x \rangle_g^\text{bare} = 0.841(28)$.

To obtain the renormalized gluon momentum fraction, we choose to renormalize nonperturbatively on the lattice using the regularization-independent momentum-subtraction (RI/MOM) scheme and then convert to the modified minimal subtraction $\overline{\text{MS}}$ scheme.
In general, the renormalized gluon moment mixes with the quark moments in the $\overline{\text{MS}}$ scheme via
\begin{equation} \label{eq:bare-moment-renormalization}
    \langle x \rangle_g^{\overline{\text{MS}}} = Z_{gg}^{\overline{\text{MS}}} (\mu^2, \mu_R^2) \langle x \rangle_g^\text{bare} + Z_{gq}^{\overline{\text{MS}}} (\mu^2, \mu_R^2) \langle x \rangle_q^\text{bare},
\end{equation}
where $Z_{gg}^{\overline{\text{MS}}}$ and $Z_{gq}^{\overline{\text{MS}}}$ are renormalization factors and the renormalization scales are $\mu$ and $\mu_R$ for the $\overline{\text{MS}}$ and RI/MOM schemes, respectively.
We use the renormalization constant $\left( Z_{gg}^{\overline{\text{MS}}}\right)^{-1} = 1.512(65)$ calculated previously in Ref.~\cite{Fan:2022qve}.
The bare quark momentum fraction, requires expensive calculations of quark disconnected diagrams, so we ignore $Z_{gq}^{\overline{\text{MS}}}$ for this calculation and introduce a 10\% systematic error on our final moment, based on previous works which report mixing as low as 2\% and no higher than 20\%~\cite{ExtendedTwistedMass:2021rdx,Alexandrou:2016ekb,Hackett:2023nkr}.
In particular, Ref.~\cite{Hackett:2023nkr} used a similar clover fermion action and found the mixing to be no more than 10\%.

We determine the renormalized gluon momentum fraction in the $\overline{\text{MS}}$ scheme at $\mu = 2$~GeV to be $\langle x \rangle_g^{\overline{\text{MS}}} = 0.557(18)_\text{stat}(24)_\text{NPR}(56)_\text{mixing}$.
We separate the statistical and nonperturbative renormalization (NPR) errors, as the NPR used a smaller number of configurations than our bare matrix-element extraction.
Our result is within two standard deviations but higher when compared with another recent lattice-QCD determination from ETMC~\cite{ExtendedTwistedMass:2024kjf}, $\langle x \rangle_g^{\overline{\text{MS}}} = 0.42(7)$, where the error is totally statistical.
The ETMC study was done at the physical pion mass and extrapolated to the continuum limit using three lattice spacings, while our study is done at a single lattice spacing and heavier-than-physical pion mass, which could partially explain the difference between the two results.
An example of a phenomenological calculation using the Dyson-Schwinger equation (DSE)~\cite{Cui:2020tdf} gives a similar result $\langle x \rangle_g^{\overline{\text{MS}}} = 0.44(2)$, whose uncertainty is estimated by varying the hadronic renormalization scale by $\pm 10\%$, but only agrees with ours within $3\sigma$.

We compare the kaon momentum fraction with the pion one, taken from previous MSULat work~\cite{Good:2023ecp}: $\langle x \rangle_g^\pi = 0.290(25)_\text{stat}(13)_\text{NPR}$, determined on the same ensemble.
We find that the gluon in the kaon carries significantly more momentum than in the pion;
their momentum faction ratio $\langle x \rangle_g^K / \langle x \rangle_g^\pi = 1.92(18)_\text{stat}$. 
This suggests that the quarks carry more momentum in the pion than the kaon, using the momentum sum rule.
This is in some tension with the findings from both the lattice study from ETMC~\cite{ExtendedTwistedMass:2024kjf} and the results from the DSE~\cite{Cui:2020tdf}, which suggest that the pion and kaon gluon moments are nearly identical.
It will be interesting to see whether a future work with physical-continuum extrapolation and full treatment of the quark mixing brings our result towards the other two results.
There is certainly hadron dependence in the behavior of the physical-continuum extrapolations.
In particular, the physical-continuum extrapolation, significantly increased the pion moment from the a12m310 ensemble result in Ref.~\cite{Good:2023ecp}, while Ref.~\cite{Fan:2022qve} found much less lattice spacing and pion mass dependence in the nucleon gluon moment.
ETMC~\cite{ExtendedTwistedMass:2024kjf} finds some differing behavior between their pion and kaon continuum extrapolations and also sees a fair bit of fluctuation from ensemble to ensemble for some of their moments, so understanding the behavior across different ensembles will be very important to fully contextualize our results.

\subsection{PDF results}
We determine the kaon gluon PDF using the pseudo-PDF method, which relates the RpITD to the lightcone gluon PDF, $g(x, \mu^2)$, through the matching relationship~\cite{MorrisIII:2022fav}
\begin{equation} \label{eq:glue-glue-matching-relation}
    \mathcal{M}(\nu, z^2) = \int_0^1 dx \; \frac{x g(x, \mu^2)}{\langle x \rangle_g} R_{gg}(x\nu, z^2\mu^2),
\end{equation}
where $\mu$ is the renormalization scale in the $\overline{\text{MS}}$ scheme and $\langle x \rangle_g^K = \int_0^1 dx \; x g(x, \mu^2)$ denotes the gluon momentum fraction in the kaon.
Although the full matching relation in Ref.~\cite{MorrisIII:2022fav} includes contributions from quark-gluon mixing through an additional kernel, $R_{gq}(x\nu, z^2\mu^2)$, in this analysis we restrict our focus to the purely gluonic term involving the gluon-gluon kernel $R_{gg}(x\nu, z^2 \mu^2)$ and neglect the quark singlet contribution.
This approximation is supported by previous gluon PDF studies~\cite{Fan:2021bcr,Fan:2022kcb,Delmar:2023agv}, which find that the systematic impact of quark contributions on the gluon PDF is significantly smaller than the current statistical uncertainties.

To extract the kaon gluon PDF, we fit the RpITD through the matching relationship defined in Eq.~\ref{eq:glue-glue-matching-relation} with an adopted phenomenologically motivated model for the PDF which has previously been used in global fits for the pion gluon PDF by JAM~\cite{Barry:2018ort,Barry:2021osv}:
\begin{equation} \label{eq:PDF-fit-form}
    f_g(x, \mu) = \frac{xg(x, \mu)}{\langle x \rangle_g (\mu)} = \frac{x^\alpha (1-x)^\beta}{B(\alpha + 1, \beta + 1)}, 
\end{equation}
where $B(\alpha + 1, \beta + 1) = \int_0^1 dx \; x^\alpha (1-x)^\beta$ is the Euler beta function, which ensures the area is normalized to unity.
Since the RpITD data are correlated, we extract the $\alpha$ and $\beta$ parameters by minimizing the $\chi^2$ in the fit
\begin{equation}\label{eq:RpITD-chi2-minimization}
    \chi^2
    = (\mathbf{d}-\mathbf{t}(\mathbf{a}))^T \mathbf{\Sigma}^{-1}(\mathbf{d}-\mathbf{t}(\mathbf{a})),
\end{equation}
where $\mathbf{d}$ is the vector of lattice-computed RpITD data, and $\mathbf{t}(\mathbf{a})$ is the vector of theoretical reconstructions at the same kinematic points, evaluated using the model in Eq.~\ref{eq:PDF-fit-form} with parameters $\mathbf{a} = (\alpha, \beta)$.
The $32 \times 32$ covariance matrix $\mathbf{\Sigma}$ encodes the statistical correlations among all data points:
\begin{equation}
    \Sigma_{ij}
    = \text{E}\left[(d_i - \text{E}[d_i])(d_j - \text{E}[d_j])\right],
\end{equation}
where $d_i$ and $d_j$ denote the RpITD values at different kinematic points, and $\text{E}[\ldots]$ is the expectation value over the jackknife samples of the data.
We perform a fit to the data with spatial separation $z \in [1,8]$ and momentum $P_z \in [2,5]$.
We exclude $P_z = 1$ from the fit, as these data decay rapidly, in a way which we find cannot be properly described by the matching.
In particular, we note Ref.~\cite{Balitsky:2019krf} states that the matrix elements have a kinematic factor proportional to $E_0^2$ in their continuum Lorentz decomposition.
If there is some additional contamination due to lattice or higher-twist effects, the double ratio defined in Eq.~\ref{eq:RpITD-double-ratio} could have a contamination which is suppressed by hadron mass and momentum, hence the lowest $P_z$ data of the mesons are less reliable.
We refer readers to the left-most columns of Figs.~7 and 8 in Ref.~\cite{Good:2024iur} to see this effect compared between the pion and nucleon.
At this level of statistics, it seems that this effect cannot be ignored, but a deeper exploration is beyond the scope of this paper, so we simply exclude the $P_z = 1$ data from the fit.

To constrain the fit in regions where the data provide limited sensitivity, we incorporate Bayesian priors on the model parameters $\alpha$ and $\beta$.
These priors are taken to be Gaussian, reflecting our prior knowledge of the expected behavior of the gluon distribution near the endpoints of $x \in [0,1]$ and enter the total negative log-likelihood as
\begin{equation} \label{eq:priors}
    \chi^2_\text{priors}
    = \frac{(\alpha - \bar{\alpha})^2}{2 \sigma_\alpha^2} + \frac{(\beta - \bar{\beta})^2}{2 \sigma_\beta^2},
\end{equation}
corresponding to independent normal distributions centered at $\bar{\alpha}$ and $\bar{\beta}$ with standard deviations $\sigma_\alpha$ and $\sigma_\beta$, respectively.
The prior means are chosen to be $( \bar{\alpha}, \bar{\beta} ) = (-0.5, 5)$, which correspond to the centers of the expected parameter ranges $\alpha \in \left[-1,0\right]$ and $\beta \in \left[0,10\right]$.
These values are consistent with global fits to meson PDFs and ensure that the functional form remains physical and well-behaved across the entire $x$-range.
The chosen prior widths, $\sigma_\alpha = \sigma_\beta = 5$, provide sufficient flexibility in the fit while guiding the parameters toward phenomenologically reasonable values.

\begin{figure*}
\centering
\includegraphics[width=0.45\textwidth]{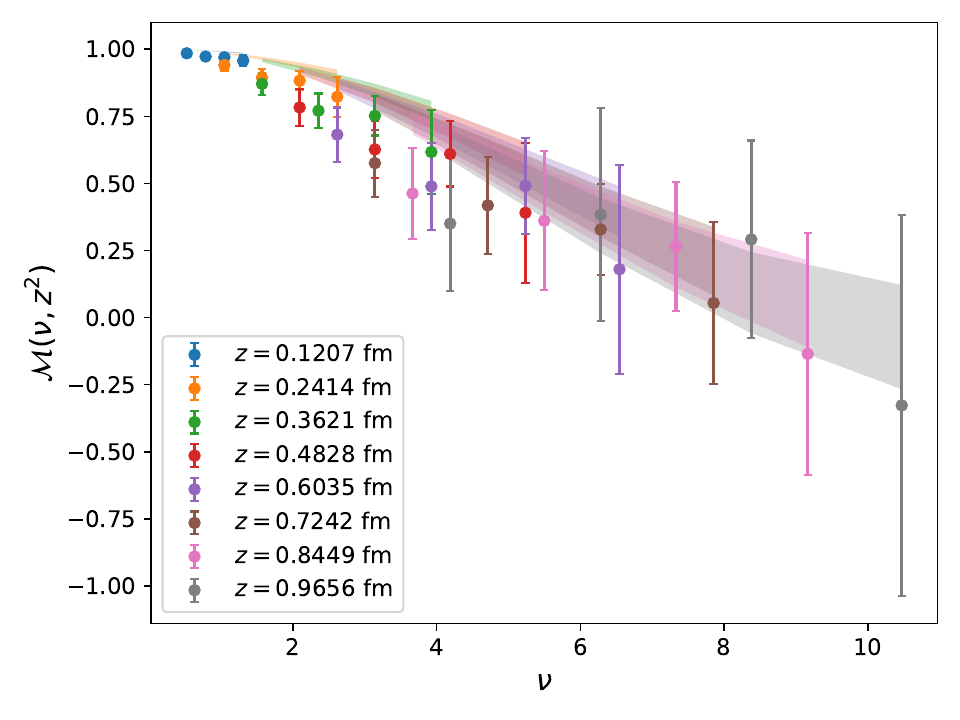}
\includegraphics[width=0.45\textwidth]{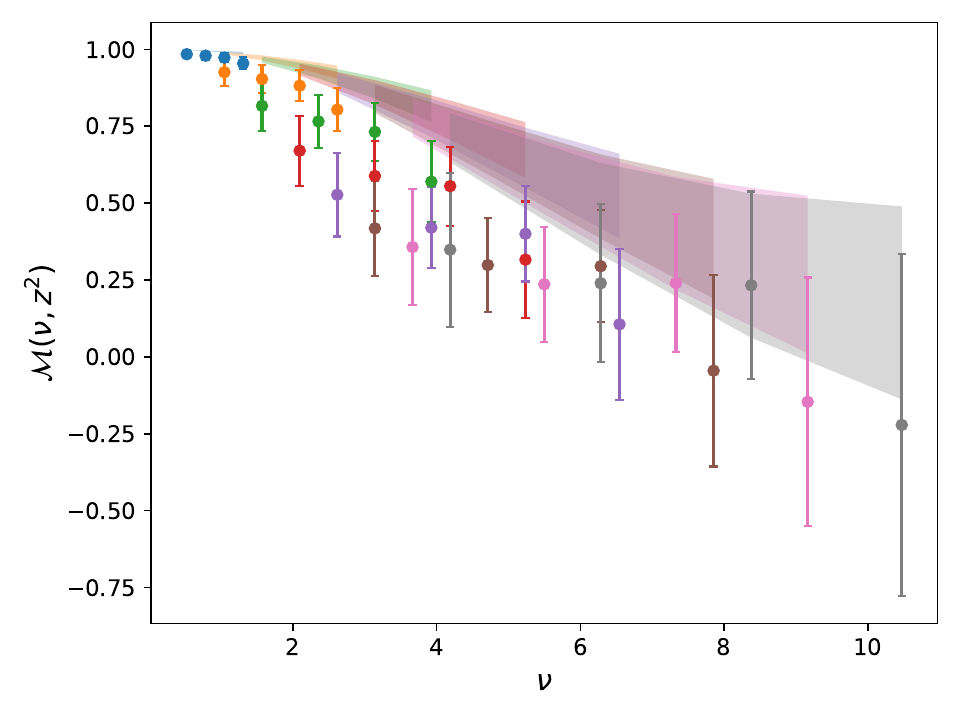}
\caption{\label{fig:RpITD_fitbands}
    Kaon (left) and pion (right) RpITD for the a12m310 ensemble.
    The bands are the gluon PDF fits from simultaneously fitting to all $z$ by minimizing the $\chi^2$ defined in Eq.~\ref{eq:RpITD-chi2-minimization} to obtain the gluon PDFs.}
\end{figure*}

The left panel of Fig.~\ref{fig:RpITD_fitbands} illustrates the fitted RpITD as a function of Ioffe-time for the Kaon, computed on the a12m310 ensemble.
Each data point corresponds to a different spatial separation $z \in [1,8]$ and momentum $P_z \in [2,5]$, while the colored bands represent the reconstructed RpITDs obtained from the fit across all $z$-values by minimizing the $\chi^2$ defined in Eq.~\ref{eq:RpITD-chi2-minimization}.
We see agreement within about $2\sigma$ between many of the data points and the reconstructed fit bands, with the smaller-$\nu$ points having less agreement with the fits.
This could be due to the aforementioned contamination of the data at small $P_z$, which has only appeared at this level of statistics.
However, it is important to appreciate that the high correlations among the data mean that the small error bars in the small-$\nu$ data do not accurately represent the true uncertainty.
The fit begins to reach $1\sigma$ agreement by around $\nu = 4$, where the diagonal of the covariance matrix (the uncorrelated error) is larger.
We find that the high statistics illuminate possible small-$P_z$ contamination and importance of the correlations, neither of which had been fully identified in previous studies at lower statistics.

\begin{figure*}
\centering
\includegraphics[width=0.45\textwidth]{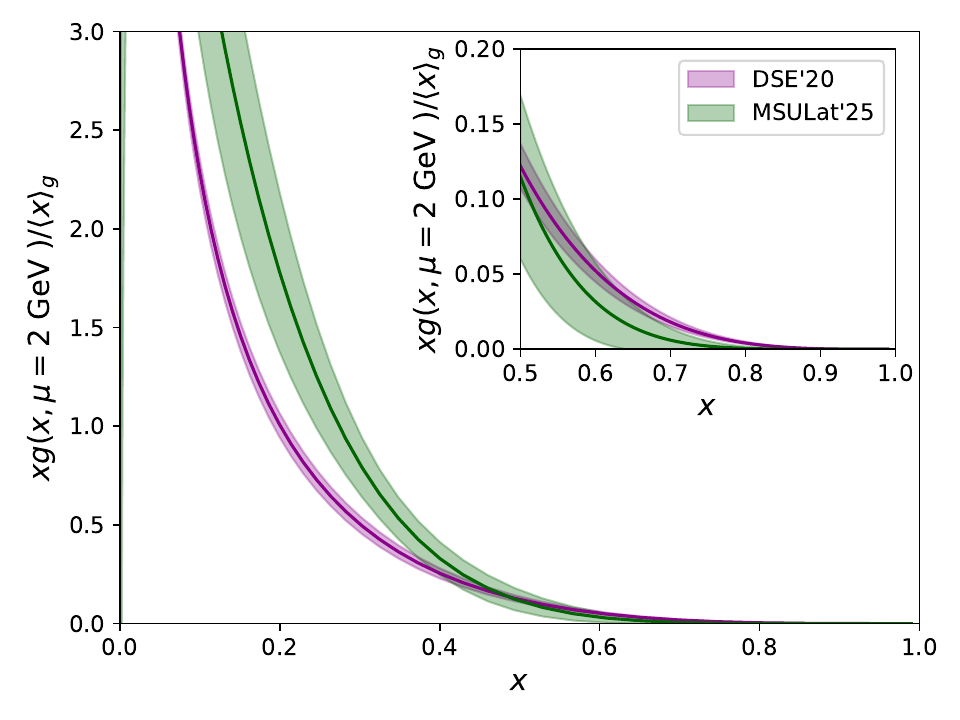}
\includegraphics[width=0.45\textwidth]{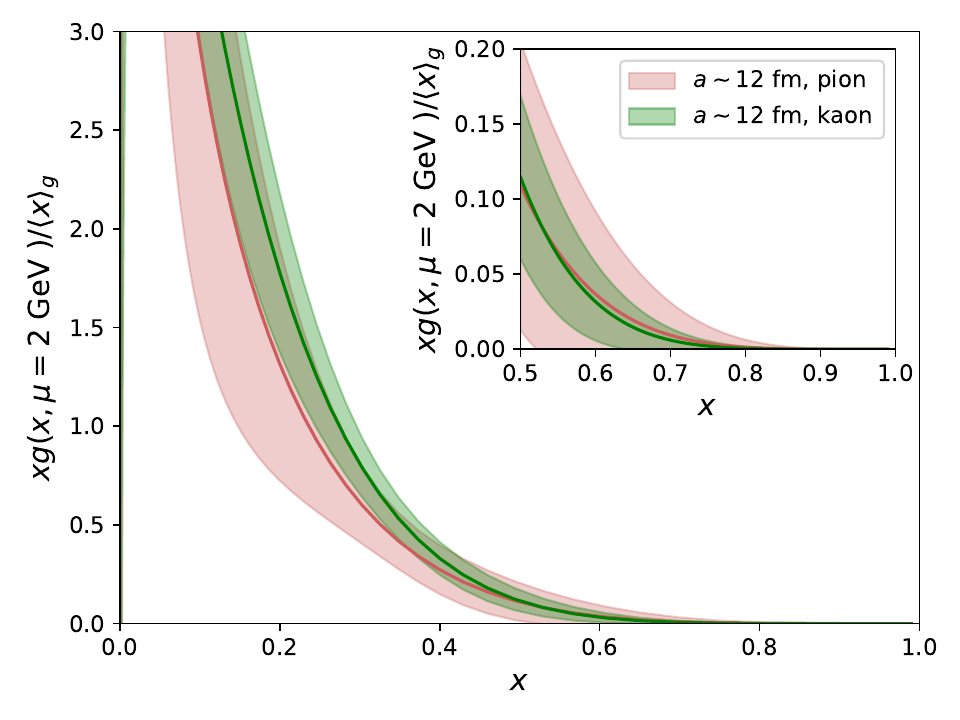}
\caption{
\label{fig:xgx}
(Left) The kaon gluon PDF $xg(x, \mu)/\langle x \rangle_g (\mu)$ as a function of $x$ obtained from the fit to the lattice data on ensembles with pion masses $M_\pi\approx 310$~MeV at $a\approx 0.12$~fm (MSULat'25), compared with the kaon gluon PDF from DSE at $\mu=2$~GeV in the $\overline{\text{MS}}$ scheme.
(Right) Comparison of pion and kaon gluon PDF $xg(x, \mu)/\langle x \rangle_g (\mu)$ as a function of $x$ with lattice spacing $a\approx 0.12$~fm pion masses $M_\pi\approx 310$~MeV, at $\mu=2$~GeV in the $\overline{\text{MS}}$ scheme.
}
\end{figure*}

Figure~\ref{fig:xgx} shows the gluon PDFs extracted from the lattice-fitted parameters $\alpha$ and $\beta$ using the normalized PDF model defined in Eq.~\ref{eq:PDF-fit-form}.
In the left panel, the kaon gluon PDF from our lattice analysis, shown in green with a shaded error band reflecting the statistical uncertainties from the RpITD fits over each jackknife sample, is compared to the DSE prediction from Ref.~\cite{Cui:2020tdf}, plotted in purple, both evaluated at $\mu = 2$ GeV in the $\overline{\text{MS}}$ scheme.
The two kaon PDFs agree comfortably within $1\sigma$ in the intermediate-$x$ regions where the lattice data has more constraining power.
Above $x\approx 0.7$, the inset shows that the PDF extracted from the lattice is slightly suppressed compared to that from the DSE, but the agreement is only just outside of $1\sigma$.
The two predictions begin to diverge below $x \approx 0.4$, likely due to the lower constraining power from the lattice data where the RpITD statistical errors getting larger with large $z$ and $P_z$ and the differences between the systematics in the two methodologies.
Removing the lattice systematics with a physical-continuum extrapolation will help better understand the systematic differences in both methodologies.

Additionally, we compare our kaon gluon PDF with that of the pion, using the matrix elements from our previous work~\cite{Good:2024iur,Good:2023ecp}.
The right panel of Fig.~\ref{fig:RpITD_fitbands} illustrates the fitted RpITD as a function of Ioffe-time for the pion, which has been fit over the same ranges of spatial separation $z \in [1,8]$ and momentum $P_z \in [2,5]$ as for the kaon.
Similar agreement is seen between the data points and the reconstructed fit bands of the pion as was previously discussed for the kaon, with the smaller-$\nu$ points having less agreement with the fits.
Again, we note that the fit begins to reach $1\sigma$ agreement near $\nu = 4$ due to the larger diagonal of the covariance matrix.

In the right panel of Fig.~\ref{fig:xgx}, we present a direct comparison between the gluon PDFs of the kaon (green) and pion (red), both extracted from the same lattice ensemble with lattice spacing $a \approx 0.12$ fm and pion mass $M_\pi \approx 310$~MeV.
The kaon PDF is more precisely determined, as indicated by its narrower error band, while the pion PDF exhibits significantly larger uncertainties across the entire $x$ range.
These broader errors stem from the noisier RpITD data for the pion, which reduce the effectiveness of the fit and limit our ability to tightly constrain its gluon distribution.
The pion and kaon PDFs have very similar shape, when the first moment is divided out, with agreement within $1\sigma$ across the entire plotted region.
By about $x \gtrsim 0.5$, the central values are almost identical, suggesting that the gluons behave very similarly in the valence region for both mesons.
At smaller $x$, we have less constraining power from the lattice data.
Overall, it will be very interesting to understand this behavior and see the effects of physical-continuum extractions of both PDFs to reduce the lattice systematics.

In addition to the kaon gluon momentum fraction, we make predictions for its normalized higher moments using the jackknife ensemble of fitted parameters $(\alpha, \beta)$ obtained from fitting the RpITD data.
Given the model PDF defined in Eq.~\ref{eq:PDF-fit-form}, the $n$th moment can be written analytically in terms of beta functions:
\begin{equation} \label{eq:moment-definition}
    \langle x^n \rangle_g
    = \int_0^1 dx \; x^n g(x)
    = \frac{B(\alpha + n, \beta + 1)}{B(\alpha+1, \beta+1)}.
\end{equation}
Using this expression, we find $\langle x^2 \rangle_g^K / \langle x \rangle_g^K = 0.123(16)$ and $\langle x^3 \rangle_g^K / \langle x \rangle_g^K = 0.0277(46)$. 
To compare with DSE, we perform a numerical quadrature on the kaon gluon PDF parameterization provided by Ref.~\cite{Cui:2020tdf} and shown in the left panel of Fig.~\ref{fig:xgx} (purple band). 
We find the DSE results to be $\langle x^2 \rangle_g^K / \langle x \rangle_g^K =  0.087(6)$ and $\langle x^3 \rangle_g^K / \langle x \rangle_g^K = 0.021(2)$, where the error is quantified using the upper and lower curves of the PDF parameterization.
Our findings deviate from these DSE results by approximately $2.1\sigma$ for $\langle x^2 \rangle_g^K / \langle x \rangle_g^K$ and $1.3 \sigma$ for $\langle x^3 \rangle_g^K / \langle x \rangle_g^K$.

\begin{figure}
\centering
\includegraphics[width=0.45\textwidth]{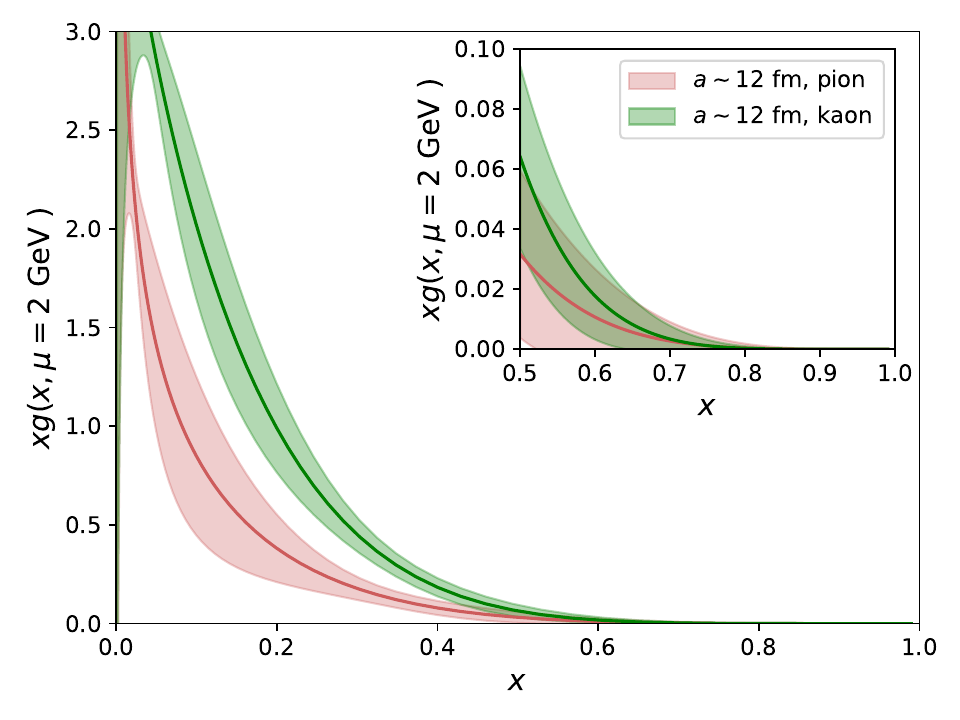}
\caption{\label{fig:FinalGluonPDF}
    The kaon (green) and pion (red) gluon PDF $xg(x, \mu = 2\text{ GeV})$ as a function of $x$.
    These are calculated by multiplying the mean of the values of $\langle x \rangle_g^K$ from our results and $\langle x \rangle_g^\pi$ from the a12m310 ensemble in Ref.~\cite{Good:2023ecp} by the curves for $xg(x, \mu)/\langle x \rangle_g (\mu)$ obtained in this work.
    Note that only the PDF errors and not the errors in $\langle x \rangle_g$ are represented here.
    }
\end{figure}

Figure~\ref{fig:FinalGluonPDF} compares the gluon PDFs, $x g(x, \mu = 2 \text{GeV})$ for the kaon (green) and pion (red), as extracted from our analysis after multiplying the curves, $x g(x, \mu = 2 \text{GeV}) / \langle x \rangle_g$, obtained from fitting the RpITD data, by the corresponding mean values of the gluon momentum fractions, $\langle x \rangle_g^K$ and $\langle x \rangle_g^\pi$.
For the pion, we use the value of $\langle x \rangle_g^\pi$ determined on the a12m310 ensemble from Ref.~\cite{Good:2023ecp}.
The bands shown in Fig.~\ref{fig:FinalGluonPDF} reflect the statistical uncertainties from the PDF fits only and do not include the uncertainties in the values of $\langle x \rangle_g^K$ and $\langle x \rangle_g^\pi$.
Despite the similar shapes seen in Fig.~\ref{fig:xgx}, the difference between the gluon structures of the kaon and pion becomes more clear in this figure due to the relative momentum fractions.
Again, it would be interesting to see if these results change at all under a physical-continuum extrapolation and to understand any other systematics that may be at play here in these extractions.
If the separation of the pion and kaon momentum fractions is confirmed, while the PDF shapes remain the same, the phenomenological effects could be quite intriguing and perhaps lead to stronger claims to be made about the emergent hadron mass and Higgs mechanism in relation to the gluonic and quark momentum fractions in mesons.

\section{Conclusions and Outlook} 
\label{sec:conclusion}
We have completed a high-statistics update of the kaon gluon PDF with around 4 times the statistics of our previous study~\cite{Salas-Chavira:2021wui} and including a new computation of the kaon gluon moment from lattice QCD.
We also studied in detail the effects on the matrix elements of various methods and amounts of gauge-link smearing.
We found that the choice of 5 steps of hypercubic (HYP5) smearing on the gauge links is a conservative one, which helps to improve the signal of gluonic matrix elements without dramatically altering the long-distance physics.
We calculated the nonperturbatively renormalized kaon gluon momentum fraction to be $\langle x \rangle_g^{\overline{\text{MS}}, K} = 0.557(18)_\text{stat}(24)_\text{NPR}(56)_\text{mixing}$ in the $\overline{\text{MS}}$ scheme at renormalization scale $\mu = 2$~GeV.
Our kaon gluon moment is slightly higher than a previous physical-continuum study from lattice QCD~\cite{ExtendedTwistedMass:2024kjf}, but the two are consistent within two standard deviations.
Future work to include a physical-continuum extrapolation of our kaon gluon moment would help to provide independent data points for lattice averaging on this quantity.

Using RpITD gluon matrix elements, we completed a correlated fit of the kaon gluon PDF using pseudo-PDF matching~\cite{Balitsky:2019krf}, ignoring the quark mixing contribution, as it has been shown to be small in previous studies~\cite{Fan:2021bcr,Fan:2022kcb,Delmar:2023agv}.
We find agreement within about $1$ to $1.5\sigma$ of the kaon PDF computed from the DSE~\cite{Cui:2020tdf} for $x\gtrsim 0.4$, and less agreement in the small-$x$ region, where the lattice data has less constraining power.
Additionally, we compared the kaon and pion PDFs divided by their first moment $\langle x\rangle_g$, calculated from the same ensemble, finding PDF shapes that are within statistical agreement.
However, the PDF $xg(x)$ for the pion and kaon are significantly different due to the different momentum fractions, with the $\langle x \rangle_g^K \approx 2\langle x \rangle_g^\pi$.

There is still much room to improve control over the systematic effects for the kaon gluon PDF with more computational resources.
In particular, finer lattice spacing, lighter pion mass, and a physical-continuum extrapolation can be explored.
It will be interesting to compare pseudo PDF with large-momentum effective theory, which was been made possible recently in Ref.~\cite{Good:2025daz};
this will still require more signal improvement to obtain reliable calculations of PDFs at reasonable levels of gauge-link smearing.
Overall, gluon observables are still quite noisy compared to those from connected-quark diagrams, and either methodology will benefit from future signal improvements, which will need to come increasingly from new techniques, rather than just high statistics.

\section*{Acknowledgments}
AN thanks Craig Roberts for providing the kaon and pion gluon PDFs at renormalization scale of 2 GeV from the DSE method for comparisons.
We thank the MILC Collaboration for sharing the lattices used to perform this study.
The LQCD calculations were performed using the Chroma software suite~\cite{Edwards:2004sx}.
This research used resources of the National Energy Research Scientific Computing Center, a DOE Office of Science User Facility supported by the Office of Science of the U.S. Department of Energy under Contract No. DE-AC02-05CH11231 through ERCAP;
facilities of the USQCD Collaboration, which are funded by the Office of Science of the U.S. Department of Energy,
and supported in part by Michigan State University through computational resources provided by the Institute for Cyber-Enabled Research (iCER).
The work of AN and WG is supported by partially by U.S. Department of Energy, Office of Science, under grant DE-SC0024053 ``High Energy Physics Computing Traineeship for Lattice Gauge Theory''.
The work of HL is partially supported
by the US National Science Foundation under grant PHY 1653405 ``CAREER: Constraining Parton Distribution Functions for New-Physics Searches'' and grant PHY~2209424.
\bibliographystyle{unsrt}
\bibliography{references}

\end{document}